\documentstyle[pra,aps]{revtex}
\begin{document}
\draft
\title{ Light Squeezing at the Transition to Quantum Chaos}
\author{Kirill N. Alekseev$^{1,2}$\cite{email1},
and Jan Pe\v{r}ina$^{3}$\cite{email2}}
\address{
$^1$Department of Physics,\AA bo Akademi, SF-20500
\AA bo, Finland\\
$^2$Theory of Nonlinear Processes Laboratory,
Kirensky Institute of Physics,
 Russian Academy of Sciences, Krasnoyarsk 660036, Russia \\
$^3$Department of Optics and Joint Laboratory of Optics\\
 Palack\'{y} University, 17. listopadu 50, 772 07 Olomouc,
 Czech Republic}
\maketitle
\begin{abstract}
We investigate theoretically the dynamics of squeezed state generation
in nonlinear systems possessing a transition from regular to chaotic
dynamics in the limit of a large number of photons. As an example, the
model of a kicked Kerr oscillator is considered. We show that at
the transition
to quantum chaos
 the maximum possible  degree of squeezing increases exponentially in time,
in contrast to the regular dynamics, where the degree of
squeezing increases in time only power-wise. We demonstrate the one-to-one
correspondence of the degree of squeezing and the value of the local
Lyapunov instability rate in corresponding classical chaotic system.
\end{abstract}
\section{Introduction}

In recent years, it was realized that chaos can have useful applications.
Among others we mention the diffusive ionization of a Hydrogen atom by
a microwave field \cite{1,2,3}, the dissociation of polyatomic molecules by
laser radiation due to the transition to chaos \cite{4,5}, the stabilization of
chaotic laser radiation \cite{6} by the controlling chaos method \cite{7},
 secret communications \cite{8} using synchronization of chaotic systems
\cite{9}, and chaos is even promisible in the generation of music
variations and other sequences of context-dependent symbols \cite{10}.
All applications mentioned dealt with classical chaos.
\par
Recently the main activity in chaos has been shifting to the studies
of quantum chaos,
particularities in the behavior of quantum systems that are chaotic in the
classical limit
\cite{11,12,12m,13,14,15}.
As a rule, quantum effects distort or
suppress ``useful'' manifestations of classical chaos. For example, in the
most studied system, a Hydrogen atom in a microwave field, quantum effects
suppress diffusive ionization by the mechanism of quantum localization
\cite{16} (the analog of Anderson localization in a solid state \cite{17})
or by the appearance of
``scars'' of a wave function \cite{18} in the vicinity of classical unstable
periodic orbits of the corresponding classical model of the atom \cite{19}.
The only probable useful application known to us (and in particular in the
area of quantum optics) for quantum chaos is the suggestion \cite{20}
to utilize the effect of wave function localization in a one-dimensional
quantum chaotic system
driven by a periodic external field, for the generation of electromagnetic
field in the Fock state.
\par
In this paper we discuss the possibility of squeezed states of the
light generation \cite{21,22,23,24} at the transition to quantum chaos.
We show that the maximal
possible degree of squeezing, achievable during some time interval of
measurement, for chaotic dynamics is much greater than for regular dynamics
during the same time interval. We find a direct correlation between the
degree of squeezing and the value of the local Lyapunov instability rate
in the corresponding classical chaotic system.
\par
Our consideration \cite{25} is based on three simple but general ideas:\\
(i) A free electromagnetic field is in a coherent state that is a Gaussian
wave packet. A wave packet  spreads when it  propagates through a nonlinear
medium. However, there
still exists a time interval of well-defined quantum-classical correspondence
during which the wave packet follows a path in phase space closed to the
path defined by
(semi)classical equations of motion. During this time interval, localized
wave packet demonstrates squeezing in one direction of phase space
and stretching in another direction. As a result, one of the field
quadrature components
may be observed in the squeezed state at some time moments. Due to the
presence of the local strong (exponential) instability inherent in the
underlying classical
chaotic dynamics \cite{31}, {\it the stretching and squeezing of the wave
packet for quantum chaos is much stronger than for the case of the regular
and stable
dynamics}, when the distance between two initially closed trajectories in
phase space increases in time only in a power wise way \cite{31}.\\
(ii) The time interval of the quantum-classical correspondence depends on
the number of quanta involved in the nonlinear dynamics: it has a power
dependence on the number
of quanta for regular dynamics, and this dependence is logarithmic for chaotic
dynamics \cite{32,11,15}. In spite of the time interval of the localized
wave packet
motion, when enhanced squeezing is expected, at chaotic dynamics may be very
short, this time scale seems quite observable in modern experiments on light
squeezing \cite{21,22,23,24}, where average number of photons $N$ involved
in the nonlinear interaction is as great as $10^3 - 10^{12}$. Moreover, in
the conditions of the experiments, the time of the localized wave packet
motion and the well-defined quantum-classical correspondence may be of the
same order as the time scale when dissipation and other factors restricting
squeezing could be neglected.\\
(iii) Along with stretching and squeezing, another inherent characteristic
of chaos in bounded phase space is folding. For strong chaos there are strong
and multiple foldings
of the wave packet; therefore, the time intervals of squeezing become very
short and
squeezing even becomes unstable \cite{28}. However, at the same time that
strong folding appears, the wave packet becomes delocalized
and well-defined quantum-classical correspondence, which we understand here
as the motion of a wave packet center along a classical trajectory, is broken
\cite{33}.
Therefore, if one studies the dynamics of the wave packet only on a time
scale of well-defined quantum-classical correspondence,
the squeezing could be strong and still rather stable for not very strong
chaos or even for regular trajectories lying in phase space near the chaotic
region.
\par
It is already known that light squeezing can be
increased near the bifurcation points between different dynamical regimes
\cite{34,24}. Among with well-studied parametric media \cite{34}, such
an increase of squeezing
was predicted also for the interaction of the field with 2-level atoms inside a
high-$Q$ cavity \cite{35,36}. The explanation of enhanced squeezing near
the bifurcation point has been used \cite{37} and it is very similar
to the arguments presented above
that utilize quantum chaos for large squeezing: At the transition
(bifurcation) between different dynamic regimes, there should exists some
diverging variable, and then the conjugate variable should be strongly squeezed.
In spite of the similarity, our suggestion  has two significant differences.
First,  all papers devoted to the study of the enhanced squeezing in
the vicinity of instability threshold \cite{24,34,35,36} dealt with the
integrable or near-integrable systems with {\it regular dynamics}.
Instead, we suggest the use of the transition from {\it regular} to
{\it chaotic} dynamics.
Second, from the general viewpoint, the transition to chaos and chaos itself
are frequent phenomena and, as a rule, they take place for a rather wide region
of the control parameters, in contrast
to a commonly narrow region of control parameter characterizing a bifurcation
between different regimes of regular motion.
\par
In this paper we consider the squeezed light generation by nonlinear
nonintegrable optical systems obeying the transition from regular to chaotic
dynamics in the classical limit. As for other problems of quantum chaos
\cite{11,12,12m,13,14,15}, we deal with
the semiclassical limit when a great number of quantum levels $N\gg 1$ are
involved into the dynamics. Our consideration is valid for any nondissipative
quantum system
with one and half degrees of freedom, but we  demonstrate our main results on
the enhanced squeezing for some particular model of the nonintegrable optical
system: the nonlinear oscillator periodically forced by the classical
field.
To investigate the nonlinear dynamics of the systems and the dynamics of
squeezing in the semiclassical limit, it is natural to use the cumulant
expansion technique \cite{38}
as a variant of the general $1/N$-expansion method \cite{39}.
In this paper, we use the
$1/N$-expansion method suggested in \cite{36}, which is
well-adopted for the problems of quantum optics. We show that as long as
the wave
packet is localized, its dynamics may be well described by the behavior of
the mean values
accounting for quantum corrections and the lowest-order quantum cumulants.
We demonstrate that the equations of motion for the second order cumulants
are essentially the same as those
are used in the definition of the Lyapunov exponent for the corresponding
classical system.
This fact gives us the opportunity to find direct correlation
between the degree of light squeezing and the
degree of local instability in the corresponding classical system.
We also compare our approach with the generalized Gaussian approximation
\cite{22,38}, which is used to find approximate solutions of many problems
of quantum optics \cite{40,41,42}, as well as with a similar variant of
the cumulant expansion \cite{43}, and find good agreement with minor differences.
\par
Our presentation is organized as follows. In the next section, using
$1/N$-expansion method and starting from the Heisenberg equations of motion,
we derive the
self-consistent set of equations for the mean values and second-order cumulants
describing the dynamics of quantum fluctuations in the semiclassical limit
for an arbitrary quantum system with one and half degrees of freedom. We show
that these equations coincide with the equations used for the calculation of
the classical Lyapunov exponent. In section 3, we discuss the conditions of
validity of our basic equations for regular and chaotic dynamics and
compare our approach with the
results of other semiclassical techniques used in both quantum optics and
quantum chaos studies. We consider the dynamics of light squeezing at the
transition to quantum chaos, and compare it with the dynamics of squeezing
for regular and stable motion in section 4.  We illustrate our results on
enhanced squeezing at the transition to chaos in the model of a kicked
quantum oscillator in section 5.
Our conclusions are summarized in the final section 6.
\section{Semiclassical dynamics of quantum fluctuations: General formalism}
We begin with a single-mode quantum system described by the Hamiltonian
$H(b,b^{\dag},t)$ including an explicit dependence on time, where $b$ and
$b^{\dag}$ are Bose operators
($[b,b^{\dag}]=1$) of the mode. We use normal ordering of operators.
Let our system include some large parameter $N\gg 1$.
The parameter $N$ may represent the number of quanta (photons) pumped to the
system \cite{30} or the number of degrees of freedom of a quantum
system \cite{36}. Because we are interested in semiclassical limit
$N\rightarrow\infty$,
it is useful to introduce new operators for the annihilation and creation
of photons
\begin{equation}
\label{1}
a=b/N^{1/2}, \quad a^{\dag}=b^{\dag}/N^{1/2}
\end{equation}
with the commutation relation
\begin{equation}
\label{2}
[a,a^{\dag}]=1/N.
\end{equation}
 In the classical limit ($N\rightarrow\infty$), one has two commuting
$c$-numbers. The natural quantum states for the consideration of the quantum
system in semiclassical limit are coherent states \cite{11,39}.
Thus we suppose that our quantum system is initially  in the coherent state
$|\alpha\rangle=\exp(N\alpha a^{\dag} -N\alpha^* a) |0\rangle$
corresponding to the mean number of quanta $\simeq N$. Following the
general scheme of
the $1/N$-expansion method \cite{39,36}, we rewrite the Hamiltonian $H$
in the form
\begin{equation}
\label{3}
H=N H_N(a, a^{\dag}, t),
\end{equation}
where the Hamiltonian $H_N$ generates the correct classical
equations of motion in the classical limit  $N\rightarrow\infty$.
In the classical limit, the operators
$a$, $a^{\dag}$, following (\ref{2}), may be considered as $c$-numbers of
order of unity and all quantum
corrections being of order of or less than $1/N$ could be treated
by perturbation theory.
\par
In order to illustrate the representation of Hamiltonian in the form (\ref{3}),
consider a particular model of the nonintegrable quantum system: the nonlinear
oscillator periodically driven by the classical field. In the interaction picture,
the Hamiltonian has the form ($\hbar=1$)
\begin{equation}
\label{4}
H=\Delta b^{\dag}b + \frac{\kappa}{2} b^{\dag 2}b^2 +
\varepsilon  N^{1/2} (b + b^{\dag}) F(t),
\end{equation}
where $b$ and $b^{\dag}$ describe a single mode
of the  quantum field and $\kappa$ is proportional to the
third-order nonlinear
susceptibility of a nonlinear medium. The last term in (\ref{4}) corresponds
to a coupling of the oscillator with an external classical periodically
modulated field containing a large number of photons $N\gg 1$ ($\varepsilon
$ is a coupling constant, $F(t)$ is a periodic function of time, and $\Delta$
is a detuning of the mode frequency from the carrier frequency of the
external field). Using new operators $a, a^{\dag}$ given in (\ref{1}),
the Hamiltonian
(\ref{4}) may be represented in the form (\ref{3}) with
\begin{equation}
\label{5}
H_N=\Delta a^{\dag} a + \frac{g}{2} a^{\dag 2} a^2 +
\varepsilon  (a + a^{\dag}) F(t),\quad g=\kappa N.
\end{equation}
It may be shown that in the classical limit $N\gg 1$, when
we have classical variables $\alpha$, $\alpha^*$
instead of operators
$a$, $a^{\dag}$ with
$\mid\alpha\mid \simeq 1$, the dependence
$g\simeq N$ correctly performs the time scale  of the energy oscillation for
Kerr nonlinearity  \cite{44}. We will return to the study of the nonlinear
oscillator (\ref{4}) in section 5.
\par
We now turn to the general case (\ref{3}) and derive the equation of motion
for mean
values and first-order cumulants. From the Heisenberg equations for
$a$, $a^2$ and their Hermitian conjugated equations,
we have the following equations of motion for the averages over coherent states
\begin{mathletters}
\label{6}
\begin{equation}
\label{6a}
i \frac{d z}{d t} = \langle V\rangle,
\end{equation}
\begin{equation}
\label{6b}
i \frac{d}{d t}\langle\left(\delta\alpha\right)^2\rangle =
2\langle V\delta\alpha\rangle+
\langle W\rangle,
\end{equation}
\begin{equation}
\label{6c}
i \frac{d}{d t}\langle\delta\alpha^*\delta\alpha\rangle =
-\langle V^*\delta\alpha\rangle+
\langle\delta\alpha^* V\rangle,\nonumber
\end{equation}
\end{mathletters}
where we have introduced
\begin{equation}
\label{7}
V(\alpha, \alpha^*)=\frac{\partial H_N}{\partial a^{\dag}},\quad
W(\alpha, \alpha^*)=\frac{1}{N} \frac{\partial V}{\partial a^{\dag}},
\end{equation}
and $z\equiv\langle a\rangle$,
$\delta\alpha\equiv a - z$,
$\langle\left(\delta\alpha\right)^2\rangle\equiv
\langle a^2\rangle - \langle a\rangle^2$,
$\langle\delta\alpha^*\delta\alpha\rangle\equiv
\langle a^{\dag} a\rangle - \langle a^{\dag}\rangle \langle a\rangle$.
In the derivation of equations (\ref{6}) we used the equalities
\cite{45}
\begin{equation}
\label{8}
\left[ a, M(a,a^{\dag}) \right]=\frac{1}{N} \frac{\partial M}{\partial a^{\dag}},
\quad
\left[ M(a,a^{\dag}), a^{\dag} \right]=\frac{1}{N} \frac{\partial M}{\partial a},
\end{equation}
which are valid for an arbitrary function $M$ of operators $a$, $a^{\dag}$
(\ref{2}).
\par
 The set of equations (\ref{6}) is not closed and actually is
equivalent to the infinite  dynamical hierarchy system for moments and
cumulants. To truncate it  we make the substitution
$a\rightarrow z +\delta\alpha$, where at least initially the mean $z\simeq 1$
and the quantum correction $|\delta\alpha(t=0)|\simeq N^{-1/2}\ll 1$.
We expand functions $V(\alpha,\alpha^*)$ and $W(\alpha,\alpha^*)$ around the
mean value $z\equiv\langle\alpha\rangle$,
\begin{equation}
\label{9}
V=V_z + \left(\frac{\partial V}{\partial\alpha}\right)_z
\delta\alpha +
\left(\frac{\partial V}{\partial\alpha^*}\right)_z
\delta\alpha^* +\cdots, \quad
 W=W_z + \left(\frac{\partial W}{\partial\alpha}\right)_z
\delta\alpha +
\left(\frac{\partial W}{\partial\alpha^*}\right)_z
\delta\alpha^* +\cdots,
\end{equation}
where the subscript $z$ means that the values of $V$, $W$, and their derivatives
are calculated at mean values $z$ and $z^*$. Substituting expansions
(\ref{9})
into Eqs. (\ref{6}) and taking into account the equality
\begin{equation}
\label{10}
\left\langle\frac{\partial V}{\partial\alpha}\right\rangle =
\left\langle
\frac{\partial^2 H_N}
{\partial\alpha\partial\alpha^*}\right\rangle =
\left\langle\frac{\partial V^*}{\partial\alpha^*}\right\rangle
\end{equation}
resulting from Eqs. (\ref{7}), we have in the first order of $1/N$ the
self-consistent set
of equations for mean values and first order cumulants
\begin{mathletters}
\label{11}
\begin{equation}
\label{11a}
i \frac{d}{d t} z  =
\langle V\rangle_z +
q[z,z^*,\langle\left(\delta\alpha\right)^2\rangle,
\langle\delta\alpha^*\delta\alpha\rangle],
\end{equation}
\begin{equation}
\label{11b}
i \frac{d}{d t}\langle\left(\delta\alpha\right)^2\rangle =
2\left(\frac{\partial V}{\partial\alpha}\right)_z
\langle\left(\delta\alpha\right)^2\rangle +
2\left(\frac{\partial V}{\partial\alpha^*}\right)_z
\langle\delta\alpha^*\delta\alpha\rangle + \langle W \rangle_z,
\end{equation}
\begin{equation}
\label{11c}
i \frac{d}{d t}\langle\delta\alpha^*\delta\alpha\rangle =
-\left(\frac{\partial V^*}{\partial\alpha}\right)_z
\langle\left(\delta\alpha\right)^2\rangle +
\left(\frac{\partial V}{\partial\alpha^*}\right)_z
\langle\left(\delta\alpha^*\right)^2\rangle.
\end{equation}
\end{mathletters}
The small quantum correction $q\simeq 1/N$
involved in the equation (\ref{11a}) has the form of the second
differential of  $V$ as follows
\begin{equation}
\label{12}
q=\frac{1}{2} d^2 V\mid_z =
\frac{1}{2}
\left(
\frac{\partial^2 V}{\partial\alpha^2}
\right)_z
\langle\left(\delta\alpha\right)^2\rangle +
\frac{1}{2}
\left(
\frac{\partial^2 V}{\partial{\alpha}^{*2}}
\right)_z
\langle\left(\delta\alpha^*\right)^2\rangle +
\left(
\frac{\partial^2 V}{\partial\alpha^*\partial\alpha}
\right)_z
\langle\delta\alpha^*\delta\alpha\rangle .
\end{equation}
The initial conditions for the system (\ref{11}) are
$\langle\left(\delta\alpha\right)^2\rangle(t=0)=
\langle\delta\alpha^* \delta\alpha\rangle(t=0)=0$
 and some arbitrary $z(0)$ is given, which is of the order of unity.
\par
We now turn to the classical equations of motion. The classical limit may be
obtained from (\ref{11a}) by neglecting the quantum correction $q$ of order
$1/N$ and then the classical Hamiltonian equations are
\begin{equation}
\label{12'}
i\frac{dz}{dt}=\frac{\partial H_N}{\partial z^*}\equiv V(z,z^*), \quad
i\frac{dz^*}{dt}=-\frac{\partial H_N}{\partial z}\equiv -V^*(z,z^*).
\end{equation}
Linearization of classical equations (\ref{12'}) near $z$  by
means of the substitution
$z\rightarrow z+\Delta\alpha$ $(|\Delta\alpha|\ll |z|$)
gives
\begin{equation}
\label{14}
i \frac{d}{d t}\Delta\alpha =
\frac{\partial V}{\partial z}
\Delta\alpha +
\frac{\partial V}{\partial z^*}
\Delta\alpha^*,\quad
i \frac{d}{d t}\Delta\alpha^* =
-\frac{\partial V^*}{\partial z}
\Delta\alpha -
\frac{\partial V^*}{\partial z^*}
\Delta\alpha^*,
\end{equation}
where all derivatives are taken on the classical trajectory
found from the Hamiltonian equations (\ref{12'}).
Using the classical analog of Eq. (\ref{10}),
$$
\frac{\partial V}{\partial z} =
\frac{\partial V^*}{\partial z^*},
$$
we have from (\ref{14}) the following equations of motion
for {\it quadratic} variables
$(\Delta\alpha)^2$ and $|\Delta\alpha|^2$
\begin{mathletters}
\label{15}
\begin{equation}
i \frac{d}{d t}\left(\Delta\alpha\right)^2 =
2\frac{\partial V}{\partial z}
\left(\Delta\alpha\right)^2 +
2\frac{\partial V}{\partial z^*}
|\Delta\alpha|^2,
\end{equation}
\begin{equation}
\label{15b}
i \frac{d}{d t}|\Delta\alpha|^2=
-\frac{\partial V^*}{\partial z}
\left(\Delta\alpha\right)^2 +
\frac{\partial V}{\partial z^*}
\left(\Delta\alpha^*\right)^2,
\end{equation}
\end{mathletters}
with the initial conditions
\begin{equation}
\label{16}
(\Delta\alpha)^2(0)=[\Delta\alpha(0)]^2, \quad
|\Delta\alpha|^2(0)=[\Delta\alpha(0)]\times [\Delta\alpha^*(0)],
\end{equation}
where $\Delta\alpha(0)$ and $\Delta\alpha^*(0)$ are the initial
values of small deviations from the classical trajectory.
We define the distance $D_{cl}$ in classical phase space between two
initially closed trajectories as
\begin{equation}
\label{17}
D_{cl}(t)=|\Delta\alpha(t)|=
\left[\left( {\rm Re}\Delta\alpha\right)^2+
\left( {\rm Im}\Delta\alpha\right)^2
\right]^{1/2},
\end{equation}
where the time-dependent quantity $|\Delta\alpha(t)|$
is determined from linearized classical
equations of motion (\ref{14}) or from {\it equivalent} equations of motion
for the square of linear deviations (\ref{15}).
The value $D_{cl}$ characterizes the degree of local instability in
the classical system: For chaotic motion $D_{cl}(t)$ increases
on average exponentially with time and
for regular motion $D_{cl}$ has only a power wise time-dependence
\cite{31}.
Using $D_{cl}(t)$, one can
define the largest Lyapunov exponent as
\begin{equation}
\label{18}
\lambda=\lim_{t\rightarrow\infty}\frac{\ln D_{cl}(t)}{t}.
\end{equation}
If $\lambda > 0$, the motion is chaotic and  $\lambda=0$ for regular
motion \cite{31}.
\par
We turn again to the quantum dynamics.
It is easy to see that the equations of motion (\ref{11b}) and (\ref{11c})
for the cumulants are the same as the classical equations (\ref{15}), except the
for the appearance of the nonlinear function $W(z,z^*)$ in Eq. (\ref{11b}),
which makes the
cumulant equations (\ref{11b}) and (\ref{11c}) nonlinear.
However, this difference
may be overcomed by introducing the new variables
\begin{equation}
\label{19}
B=N\langle\delta\alpha^*\delta\alpha\rangle+\frac{1}{2},\quad
C=N\langle\left(\delta\alpha\right)^2\rangle,
\end{equation}
where $B$ is real and $C$ is complex.
Using (\ref{7}), the quantum equations of motion (\ref{11}) may be
rewritten in variables (\ref{19}) in the following form
\begin{mathletters}
\label{20}
\begin{equation}
\label{20a}
i \dot{z} =
\langle V\rangle_z + \frac{1}{N}Q(z,z^*,B,C,C^*),
\end{equation}
\begin{equation}
\label{20b}
Q(z,z^*,B,C,C^*)=\frac{1}{2}\left(\frac{\partial^2 V}{\partial\alpha^2}
\right)_z C +
\frac{1}{2} \left(\frac{\partial^2 V}{\partial\alpha^{*2}}
\right)_z C^* +
\left(\frac{\partial^2 V}{\partial\alpha^{*}\partial\alpha}\right)_z
\left(B-\frac{1}{2}\right),
\end{equation}
\begin{equation}
\label{20c}
i\dot{C}=2\left(\frac{\partial V}{\partial\alpha}\right)_z C +
2\left(\frac{\partial V}{\partial\alpha^*}\right)_z B,
\end{equation}
\begin{equation}
\label{20d}
i\dot{B}=-\left(\frac{\partial V^*}{\partial\alpha}\right)_z C +
\left(\frac{\partial V}{\partial\alpha^*}\right)_z C^*
\end{equation}
\end{mathletters}
and the corresponding equation for $C^*(t)$ that could be obtained from
(\ref{20c}) by complex conjugation.
Initial conditions for the system (20) are
\begin{equation}
\label{21}
B(0)=1/2,\quad C(0)=C^*(0)=0.
\end{equation}
Now all variables are of the order of unity and the dependence on the small
parameter $1/N$ is present explicitly only in the expression for the
quantum correction to classical motion in (\ref{20a}).
\par
Compare the set of classical Hamiltonian equations (\ref{12'}) and
equations of motion (\ref{15}) for classical linear fluctuations with
the equations of motion (\ref{20}) for mean values and quantum cumulants.
It is evident that both sets of equations have the same structure with
the following principal differences. First, the quantum equations
(\ref{20c}) and (\ref{20d}) for the cumulants, in contrast
to the classical equations (\ref{15}),
are calculated near the mean value $z$ and take into account the quantum
correction $Q$ [Eqs. (\ref{20a}), (\ref{20b})]. Second, it should be noticed
that it is impossible
to obtain the initial conditions (\ref{21}) for $C$ and $B$
from the initial conditions for the classical equations (\ref{16}).
However, if one considers the case of large $N$, when the quantum correction is
small, the quantum equations (\ref{20}) are {\it identical to the
classical equations
(\ref{15}) used in the definition of the maximum Lyapunov exponent}.
The same conclusion on the equivalence of the equations of motion
for low-order cumulants
and equations of motion arising in the definition of the Lyapunov
exponent has been
obtained earlier for the generalized Tavis-Cammings model in \cite{28}
and for systems with Hamiltonians consisting of the sum of kinetic and
potential energies in \cite{43}.
\par
The self-consistent system of equations (\ref{20}) completely describes the
dynamics of quantum fluctuations in the first order of $1/N$. We shall use
these equations for the description of the dynamics of light squeezing at
the transition
to quantum chaos in section 4 after a discussion of the validity of our
approach and its comparison with other semiclassical methods.
\section{Conditions of validity and comparison with other approaches}
Here we discuss the condition of validity of the $1/N$-expansion (\ref{9})
and our equations
of motion (\ref{20}) for the mean values and cumulants. In analogy to
the classical
distance (\ref{17}), we introduce the ``distance'' $D_q$ for the quantum case as
\begin{equation}
\label{22}
D_q(t)=\frac{1}{N} B^{1/2}(t)
\simeq\left[\langle|\delta\alpha|^2\rangle\right]^{1/2}.
\end{equation}
It is easy to see that $D_q$ coincides with the ``convergence radius''
$r=[{\rm Re} (\delta\alpha)]^2+[{\rm Im} (\delta\alpha)]^2$ of the
$1/N$-expansion (\ref{9}).
Initially $D_q(0)\simeq 1/N\ll 1$, and if  $D_q(t)\ll 1$ during some time
interval, then the $1/N$-expansion is well-defined in this time interval.
\par
When quantum correction in Eq. (\ref{20a}) is small, equations (\ref{20}) are
identical to the classical equations (\ref{15}) arising in the definition of
the classical Lyapunov exponent (\ref{18}). Thus, for classically chaotic motion,
we have exponential growth of $B(t)$, $C(t)$, and $D_q(t)$ as
\begin{equation}
\label{23}
D_q(t)\simeq D_q(0)\exp(\lambda\omega_0 t),
\end{equation}
where $\lambda$ is the Lyapunov exponent and $\omega_0$ is the characteristic
frequency in the classical dynamic system. For example, $\omega_0=g$ in the model
of the Kerr oscillator (\ref{5}).
From the conditions $D_q(t)\ll 1$ and $D_q(0)\simeq 1/N$,
we have the following estimate for the time scale of validity of
our semiclassical approach
\begin{equation}
\label{24}
t\ll t^*=\frac{1}{\lambda\omega_0}\ln N.
\end{equation}
In contrast, for classically regular motion $D_q(t)$ increases with time only
power wise: $D_q(t)\simeq t^{\gamma}$, where the index $\gamma\simeq 1$
depends on the system under study. For example, $\gamma=1$ for
nonlinear oscillator with the Hamiltonian (\ref{5}) for $\varepsilon=0$, as well as
for integrable nonlinear oscillators with power wise higher-order
nonlinearity \cite{45''}.
As a result, the time
scale of validity of our approach in the case of regular motion is
\begin{equation}
\label{25}
t\ll t^*={\omega_0}^{-1} N^{1/\gamma}
\end{equation}
and it is much greater than (\ref{24}) for chaotic motion.
The time scale $t^*$ has very simple physical meaning: during time interval
$t^*$ we have localized wave packet with center moving along the classical
path.
\par
The scale (\ref{24}) of the well-defined
quantum-classical correspondence for chaotic systems
was first introduced in \cite{32}, and now it is investigated in
details in different systems (for a review, see \cite{15}).
The time scale (\ref{24}) is very short for small $N$. However, under
the conditions of
modern experiments on light squeezing, where the average number of
photons $N$ involved in the nonlinear interaction is $10^3-10^{12}$, $t^*$ is of the order of
$10-100$ of the system's characteristic periods $\omega_0^{-1}$
and thus it looks quite reasonable.
\par
We now compare our approach to the description of the dynamics of quantum
fluctuations with that widely used in quantum optics and called
the generalized Gaussian
approximation \cite{22,38,40,41,42}. This approximation also assumes the
existence of a well-localized, almost Gaussian wave packet throughout the
time of quantum evolution and in this respect it is very closed to our consideration.
More precisely, the generalized Gaussian approximation consists in an assumption
that the Fourier transform of the quantum distribution function, i.e.
the quantum characteristic function, is Gaussian for any moment of time
\cite{38}. Such a quantum state corresponds to the superposition of the
coherent signal and small quantum noise \cite{22}. For such a state,
only the first- and second-order cumulants are nonzero. It possesses
the expression for
higher-order cumulants in terms of only the first and
second-order cumulants and
truncation of an infinite hierarchy dynamic system for cumulants is possible.
\par
We compare our system (\ref{20}) with the dynamic
equations for the mean values and cumulants
obtained within the generalized Gaussian approximation for several popular
models of quantum optics: second harmonic generation described by the
Hamiltonian
\begin{equation}
\label{25'}
H=\omega_1 b_1^{\dag}b_1 +\omega_2 b_2^{\dag}b_2 -
\left(\kappa b_1^{2}b_2^{\dag} + {\rm H.c.}\right),
\quad [b,b^{\dag}]=1,
\quad \omega_2=2\omega_1,
\end{equation}
the problem of nondegenerate optical 3-wave mixing described by the Hamiltonian
\begin{equation}
\label{25''}
H=\sum_{j=1}^{3} \omega_j b_j^{\dag}b_j  -
\left( \kappa b_1 b_2 b_3^{\dag} + {\rm H.c.}\right),
\quad \omega_3=\omega_1+\omega_2,
\end{equation}
and for the periodically forced nonlinear oscillator with the Hamiltonian (\ref{4}).
Considerations of these problems within the generalized Gaussian approximation
are presented in Refs. \cite{40}, \cite{41}, and \cite{42},
respectively. For the forced nonlinear oscillator, we found that our
self-consistent set of
equations (\ref{11}) and (\ref{12}) coincides with the corresponding basic equations
of \cite{42} up to terms of the order of $1/N^2$, and for the problems of
nondegenerate and degenerate optical 3-wave mixing, our approach gives
equations that are
identical to the corresponding equations of motion obtained
within the generalized
Gaussian approximation \cite{40,41,46}.
\par
We turn now to the discussion of squeezing at the transition from regular
to chaotic motion.
\section{Squeezing at the transition to quantum chaos}
Define the general field quadrature as
$X(\theta)=a\exp(-i\theta)+a^{\dag} \exp(i\theta)$,
where $\theta$ is a local oscillator phase. A state is said to be squeezed if
there is some phase $\theta$ for which the variance of $X_{\theta}$
\begin{equation}
\label{26}
\langle\left(\delta X\right)^2\rangle =
\langle\left(\delta\alpha\right)^2\rangle \exp(-i 2\theta)+
\langle\left(\delta\alpha^*\right)^2\rangle \exp(i 2\theta)+
2 \langle|\delta\alpha|^2\rangle+1/N
\end{equation}
is less than the variance for coherent state or the vacuum \cite{21}.
In variables (\ref{19}), the condition of squeezing takes the form
\begin{equation}
\label{27}
S(\theta)=2 B+C\exp(-i 2\theta)+C^*\exp(i 2\theta)<1.
\end{equation}
The minimum of the variance of the general field quadrature (\ref{27})
with respect to
its dependence on the local oscillator phase $\theta$ is reached at
$\theta_{min}$, determined as
\begin{equation}
\label{28}
\theta_{min}=(\varphi-\pi)/2,
\end{equation}
where $\varphi$ is the argument of the cumulant $C$
[$C\equiv|C|\exp(i\varphi)$].
For $\theta=\theta_{min}$, the condition of squeezing (\ref{27}) is
\begin{equation}
\label{29}
S\equiv S(\theta_{min})=2( B-|C|)<1.
\end{equation}
The value $S$ determines the minimum half-axis of the quantum
noise ellipse \cite{22}. The condition (\ref{29}) is called
{\it principal squeezing}
because it gives the maximum squeezing measurable by
homodyne detection \cite{47,22}.
\par
Let us now compare the dynamics of the principal squeezing for classically
regular and chaotic motion. Initially the Gaussian wave packet spreads when
it propagates through a nonlinear medium. However, there still exists the time interval
of the well-defined quantum-classical correspondence (\ref{24}) or (\ref{25})
during which the wave packet's center follows a path in phase space governed by
the semiclassical equation of motion (\ref{20a}). Moreover, because our
equations of motion for cumulants (\ref{20c}) and (\ref{20d}) in fact
coincide with the
equations arising from the definition of the maximum Lyapunov exponent,
we can apply simple physical arguments to the strong deformation of
the classical phase volume
at chaos for the prediction of the strong squeezing of the noise ellipse at
quantum chaos in the semiclassical limit.
\par
Due to the presence of the strong (exponential) local instability inherent
in the underlying classical chaotic dynamics, a quantum noise ellipse
may be strongly
stretched in one direction and squeezed in another direction. As a result,
the value of principal squeezing $S$ in Eq. (\ref{29}), which correlates with
the minimum half-axis of the quantum noise ellipse, on average exponentially
decreases in time
\begin{equation}
\label{sq-ch}
S(t)\simeq \exp(-\lambda\omega_0 t).
\end{equation}
The stretching and squeezing
of a noise ellipse at quantum chaos is much stronger than for the case of
regular and stable dynamics, when the distance between two initially closed
trajectories in phase space increases in time in a power wise way
resulting in only a
power wise decrease of the principal squeezing in time
\begin{equation}
\label{sq-reg}
S(t)\simeq (\omega_0 t)^{-\beta},
\end{equation}
where the constant $\beta\simeq 1$. As an example of such time-dependence,
we consider the Kerr oscillator (\ref{5}) without an external field
($\varepsilon=0$). In this case the model is integrable, and from
an exact solution in the limit $N\gg 1$ \cite{45'}, we get
\begin{equation}
\label{example}
S(t)=1+2|\alpha_0|^2 g t \left[ |\alpha_0|^2 g t -
\left( 1+ |\alpha_0|^4 g^2 t^2 \right)^{1/2} \right]
\stackrel{g t\gg 1}{\longrightarrow}
\left( 6|\alpha_0|^2 g t \right)^{-1},
\end{equation}
where $\alpha_0$ is some initial condition
($|\alpha_0|\simeq 1$). The same result could be obtained
directly from our semiclassical approach  neglecting quantum
correction in Eq. (\ref{20}) and combining it with Eq. (\ref{29})
\cite{45''}.
\par
Making a comparison of squeezing for regular and chaotic motion,
several comments are necessary.\\
1. Both formulas (\ref{sq-ch}) and (\ref{sq-reg}) are obtained within
the pure classical picture when quantum corrections are neglected.
The deviations from dependencies (\ref{sq-ch}) and (\ref{sq-reg})
are expected when quantum corrections become sufficient,
especially at $t\simeq t^*$.\\
2. Squeezing for chaotic dynamics is exponential only on average.
Actually, the rate of phase volume deformation and, consequently,
the rate of quantum noise ellipse deformation and squeezing
are directly related in the semiclassical limit ($N\gg 1$)
to the degree of local Lyapunov instability $D_q$ [Eq. (\ref{22})] in
the system, which at some moments in time may be different from the
maximum Lyapunov exponent
(\ref{18}) measured at asymptotics $t\rightarrow\infty$.
Such behavior is typical, for instance, for a chaotic trajectory
that spends some time near a stability island. In this case,
the local instability and statistical properties of chaotic motion
are weak \cite{beloshapkin}. Finally, the trajectory escapes
to a large chaotic sea, and the local instability becomes strong.
Of course, the Lyapunov exponent is positive in both cases.
Formally, the resulting nonmonotonic
dependence of squeezing on time could be modelled by a slowly
varying dependence of the parameter $\lambda$ [Eq. (\ref{sq-ch})].
We will say more on this problem in the next section.\\
3. Along with wide class of integrable systems
with stable regular dynamics, there is a small class of systems with
regular but unstable dynamics, for which the distance $D_q$
given in (\ref{22})
increases in time exponentially. In such a case the systems
are in the state near the
bifurcation point between different dynamic behaviors and they are
exponentially unstable. The mechanism of enhanced squeezing in unstable systems with
regular dynamics is very similar to that discussed at the transition
to quantum chaos and  has been discussed in detail in \cite{34,35,36,37}.\\
4. Considering squeezing at chaos, we have not discussed yet the influence
of one of the main characteristics of chaos: folding of the phase volume
that in addition to stretching, is present in any bounded Hamiltonian system.
For strong chaos, strong and multiple folding of the phase volume appears
during the time scale of well-defined quantum-classical correspondence.
The perimeter of the
phase volume increases in time exponentially, and eventually the phase volume
envelope gets some fractal-like structure \cite{31}. Moreover, the final
shape of the phase volume is very sensitive to small changes of the initial
conditions and/or parameters. The complex and unstable evolution of the phase
volume is displayed in the time-dependence  of  squeezing \cite{28}. For strong
chaos and long enough time, squeezing becomes strongly dependent on
tiny variations of the system's parameters. Such a regime of squeezing was
called in \cite{28} {\it unstable squeezing}.
For unstable squeezing,
the range of the local oscillator phase, for which squeezing is possible, becomes
so narrow that it makes observation of squeezing practically
impossible \cite{28}.
\par
It should be noticed that at the same time when multiple folding of wave
packet appears, the quantum-classical correspondence breaks down. For strong chaos
this time scale is very short. In contrast, for weak chaos,
the time scale of the well-defined quantum-classical correspondence and
time scale of strong folding may be rather long in the semiclassical
limit resulting in stable and enchanted squeezing. The same arguments should be
valid for regular trajectories located near the chaotic motion in phase space.
Actually, during a finite time interval stability properties of a given trajectory
are determined by the time-dependence of distances (\ref{17}) or (\ref{22})
between two initially nearby trajectories. At a finite time the value of
$D_q$ for the regular trajectory is often practically indistinguishable from
the nearby chaotic trajectory \cite{47'} resulting in the same degree of squeezing.
\par
We will illustrate a general picture of squeezing at the transition to
quantum chaos through an example of the model of the kicked nonlinear oscillator in
the next section.
\par
Concluding this section, let us compare qualitatively
the maximum achievable degree of
principal squeezing for regular and chaotic motion during the time interval
of the well-defined quantum-classical correspondence. First consider the case of
squeezing at chaos. Principal squeezing on average
is exponential (\ref{sq-ch}),
and at the end of time interval (\ref{24}) it should be not
less than $S(t^*)\simeq\exp(-\lambda\omega_0 t^*)\simeq 1/N$.
For regular dynamics, the principal squeezing has
a power wise time-dependence (\ref{sq-reg})
and it should be not less than
$S(t^{*})\simeq {(\omega_0 t^*)}^{-\beta}\simeq N^{-\beta/\gamma}$,
where constants
$\gamma\simeq 1$ and  $\beta\simeq 1$, and we used the estimate
(\ref{25}) of $t^*$ for
regular motion.
\par
Thus these rough estimates show that
the degree of squeezing at time $t^*$ is
comparable in the cases of regular and chaotic dynamics,
but because $t^*$ is
much shorter for chaotic dynamics than for regular,
squeezing is much faster in the chaotic systems.
\section{Example: kicked nonlinear oscillator}
Consider a nonlinear oscillator interacting with a time-periodic field
as given in Eqs. (\ref{4}) and (\ref{5}).
This model describes, for example, a high-{\it Q}
cavity  filled by a medium with Kerr nonlinearity and excited by an external
laser field \cite{20}. The same effective Hamiltonian may also govern the
interaction of a laser field with a high-density excitons in a semiconductors
\cite{48}. The different variations of model discussed are very popular in
both quantum optics \cite{49} and quantum chaos \cite{11,42,50} studies.
\par
In this paper we choose the form of $F(t)$ in (\ref{5}) as a periodic
sequence of kicks
\begin{equation}
\label{31'}
\delta_T(t)=\sum_{n=-\infty}^{\infty}\delta(t-nT),
\end{equation}
where $\delta(t)$ is the Dirac $\delta$-function.
In an experiment, a sequence of
short light pulses can be generated by a mode locked laser.
The use of the sequence of kicks possesses to obtain discrete maps instead of
differential equations and it sufficiently simplifies computations and
reduces
numerical errors, which is especially important when we analyze the influence
of small quantum corrections ($N\gg 1$) on the dynamics of squeezing.
\par
In what follows we shall use scaled variables $t^{\prime}=g t$ and
$\Delta^{\prime}=\Delta/g$
measuring time and detuning
in the units of the coupling constant $g$, and then we omit primes,
which is formally equivalent to choosing
$g\equiv 1$ in the Hamiltonian (\ref{5}).
\par
The classical equation of motion (\ref{12'}) obtained
from the Hamiltonian (\ref{5})
has the form
\begin{equation}
\label{30}
i\frac{d z}{d t}=\Delta z + g\mid z\mid^2 z + \varepsilon\delta_T(t).
\end{equation}
From Eq. (\ref{30}) and using a standard technique \cite{11,31},
we have the classical map
\begin{equation}
\label{31}
z_{n+1}=F(z_n), \quad
F(z_n)=\exp[-i T (\Delta+|z_n-i\varepsilon|^2)](z_n-\varepsilon),
\end{equation}
where $z_n$ is the value of $z$ just before $n$-th kick, $z_n\equiv z(nT-0)$.
It may be shown analytically \cite{50} that the approximate criterion of
transition from regular motion to strong and global chaos for the map
(\ref{31}) in the limiting case $\varepsilon\ll 1$ is
\begin{equation}
\label{31a}
K=2\varepsilon T \gtrsim 1.
\end{equation}
We illustrate the dynamics governed by the map (\ref{31}) in Fig.~\ref{fig1},
where phase portraits (a,c,e) and the time dependence of intensity
$|z_n|^2$ (b,d,f) are shown for the cases of regular motion,
weak chaos, and strong chaos correspondingly.
\par
We now turn to the dynamics of quantum cumulants and the mean values with quantum
corrections. Using the smallness of $1/N$, we obtain from the
semiclassical equations
of motion (\ref{20}) and the classical map (\ref{31}) the following coupled
maps describing the dynamics of mean values and cumulants (details  of
the derivation of these maps are presented in the Appendix)
\begin{mathletters}
\label{32}
\begin{equation}
z_{n+1}=F(z_n)+\frac{1}{N}Q_n,
\label{32a}
\end{equation}
\begin{eqnarray}
 C_{n+1} & = & \exp \left[ -i 2 T
\left( \Delta+|z_n-i\varepsilon|^2\right) \right]
 \times \nonumber \\
 & & \mbox{} \biggl\{ -2 \left( z_n-i\varepsilon \right)^2
\left( T^2 |z_n-i\varepsilon |^2 + i T \right) B_n +
\left( 1-2 i T | z_n-i\varepsilon |^2\right)
 C_n  \nonumber\\
 & & \mbox{} -T^2 \left( z_n-i\varepsilon \right)^2
\left[ \left( z_n-i\varepsilon \right)^2 C^*_n +
{\rm c. c.}\right]
\biggr\} ,
\label{32b}
\end{eqnarray}
\begin{eqnarray}
B_{n+1} & = & \left( 2 T^2 |z_n-i\varepsilon |^4
+1\right) B_n+
T^2 |z_n-i\varepsilon |^2
\left[ \left( z_n-i\varepsilon\right)^2 C^*_n
+{\rm c.c.} \right] \nonumber\\
 & & + \left[ -i T \left( z_n-i\varepsilon\right)^2 C^*_n
+{\rm c.c.} \right] ,
\label{32c}
\end{eqnarray}
\end{mathletters}
where $F(z_n)$ in (\ref{32a}) is the classical map introduced in (\ref{31})
and the quantum correction $Q_n$ in (\ref{32a}) has the form
\begin{mathletters}
\label{33}
\begin{equation}
Q_n=A_1 (z_n,z_n^*) C_n + A_2 (z_n,z_n^*) C^*_n +
A_3 (z_n,z_n^*) (B_n-1/2),
\label{33a}
\end{equation}
\begin{equation}
A_1 (z_n,z_n^*)=\frac{1}{2} e^{-i T (\Delta+\mid z_n-i\varepsilon\mid^2)}
\left[-i T (z_n^*+i\varepsilon) (2-i T |z_n-i\varepsilon |^2)\right],
\label{33b}
\end{equation}
\begin{equation}
A_2 (z_n,z_n^*)=\frac{1}{2} e^{-i T (\Delta+\mid z_n-i\varepsilon\mid^2)}
\left[ -T^2 (z_n-i\varepsilon)^3\right],
\label{33c}
\end{equation}
\begin{equation}
A_3 (z_n,z_n^*)=- e^{-i T (\Delta+\mid z_n-i\varepsilon\mid^2)}
T (z_n-i\varepsilon) (T | z_n-i\varepsilon |^2+2i).
\label{33d}
\end{equation}
\end{mathletters}
Following (\ref{21}) the initial conditions for the maps (\ref{32})
are $B_0=1/2$, $C_0=0$, and arbitrary $z_0$ of order of unity.
\par
The self-consistent set of maps (\ref{32}) and (\ref{33}) determines
the dynamics of the quantum fluctuations for kicked nonlinear oscillator
in first order of $1/N$.
\par
Now we want to compare the time-dependence of the principal squeezing
$S$ in (\ref{29}),
the degree of local instability (\ref{22}), and the quantum correction $Q$
in (\ref{33}) for three characteristic cases of classical dynamics:
regular motion, mild chaos and hard chaos. To avoid the
dependence of $D_q$ on $N$ it is useful to introduce new normalized
distance
$$
d=N D_q.
$$
Phase portraits and time-dependencies of the intensity $|z|^2$
for the three characteristic cases of classical dynamics governed by the map
(\ref{31}) are shown  in Fig. 1. The time-dependences of ${\log}_{10} S$,
${\log}_{10} d$, and ${\log}_{10} Q$ for a large but finite number of quanta
$N=10^9$ are shown in Fig. 2, where curves 1, 2, and 3 correspond to the
cases of regular motion ($\lambda=0$) and chaotic motion with
an increasing value of the Lyapunov exponent $\lambda$, respectively.
As is evident from comparison of Fig. 2a
and Fig. 2b, the strongest local instability determines the highest degree
of squeezing. It should be mentioned that the difference in the magnitude
of principal squeezing for chaotic motion and regular motion is of
several orders during only several kicks. A power wise time-dependence of $d$
(Fig. 2b, curve 1) is assisted by the corresponding slow growth in time of
the  quantum
correction (Fig. 2c, curve 1). In contrast, for chaotic motion, the growth of
$d$ and $Q$ is exponential (curves 2 and 3 in Fig. 2b and 2c) resulting in
a logarithmic dependence of the applicability of semiclassical approach on the number
of quanta in (\ref{24}).
\par
The increment of local instability and the corresponding degree of squeezing
are complex functions of the initial conditions or parameters.
As a result, the dependence of squeezing on initial conditions
or parameters may be rather complex. In Fig. 3, we plot the minimum value
$\min{\log}_{10}S$ of the principal squeezing  during seven kicks and degree
of the local
instability ${\log}_{10} d$ as a function of the kick's period $T$ at fixed initial
conditions, $\varepsilon=0.1$, and number of quanta $N=10^9$. The regions of
parameter values with mainly regular or chaotic motion in the classical
limit $(N \rightarrow \infty)$ are marked in the center of the plot. The first
large region of
chaos begins at $T_{cr}\approx 5.4$, which is in good agreement with the
approximate criterion (\ref{31a}) of the transition to chaos.
Of course, there are some singular values of the parameter $T$ corresponding to regular motion in the range
of $T$ that is mainly filled by chaotic trajectories.
\par
As is evident from Fig. 3, there is a direct correlation between
the degree of
squeezing and degree of instability ${\log}_{10} d$.
Because we present the dependences of ${\log}_{10} d$
and squeezing on $T$ for only seven kicks, the degree of instability and
squeezing may be comparable for some points from regular and chaotic
ranges of $T$. However, generally, as it follows from (\ref{31a}), the degree
of instability increases with increasing $T$ and correspondingly we have
an increase of squeezing (Fig. 3).
\par
In Fig. 4 we plot ${\log}_{10} d$ and the minimum of
${\log}_{10} S$ during nine kicks as
a function of another parameter $\varepsilon$ and at fixed $T$ and initial
conditions. Again, the correlation of local instability and squeezing as well
as enhancing of squeezing at the transition to chaos are visible. The same
dependences but for three kicks instead of nine are shown in Fig. 5a.
In spite of the correlation between the degree of squeezing and $ {\log}_{10} d$
it is also evident in Fig. 5a that the increase of squeezing at the
transition from regular to chaotic motion $(\varepsilon_{cr}\approx 0.5)$
is smoothed more than in the case of nine kicks shown in Fig. 4.
\par
As we already
mentioned, the degree of local instability may itself depend on time.
This is illustrated in Fig. 5b, where in an interval of $3 - 8$ kicks the instability
for $\varepsilon=1.8$ (curve 1) is greater than for $\varepsilon=2.22$
(curve 2). As a result, the squeezing during 3 kicks at $\varepsilon=1.8$
(point 1 in Fig. 5a) is greater than for $\varepsilon=2.22$
(point 2 in Fig. 5a). After eight kicks we have inverse picture: the instability
for $\varepsilon=2.22$ is greater than at $\varepsilon=1.8$ (see Fig. 5b)
resulting in enhanced squeezing for the parameter's values of point 1
in comparison to squeezing for the parameter's values of point 2.
\par
Summarizing our findings from Figs. 3-5, we see that in the semiclassical limit
the degree of squeezing has a one-to-one correspondence to the degree of local
instability in the system. For a short time of the order of several kicks,
the dependence of squeezing on the Lyapunov exponent, which is calculated
in the limit $t\rightarrow\infty$, is not well pronounced.
Moreover, squeezing for regular motion located near the border of the transition
to chaos is comparable to squeezing for weak chaos.
The dependence of squeezing on the value of the Lyapunov exponent becomes
more pronounced for moderate times of the order of ten kicks.
\par
We compare now the time-dependence of principal squeezing for different $N$.
The dependence of $S$ on $t$ at $N \rightarrow \infty$
(curve 1) and at $N=10^7$ (curve 2) is presented in Fig. 6
for the case of mild chaos.
Curves 1 and 2 coincide during seven kicks, which determines the time interval
of well-defined quantum-classical correspondence for such a choice of parameters.
Moreover, our numerical calculations demonstrate that if the number
of quanta increases in two orders up to $N=10^9$,
the squeezing is indistinguishable from
squeezing at $N \rightarrow \infty$ up to fifteen kicks.
This finding is well explained by
remembering  $\ln{N}$ dependence (\ref{24}) of the time scale.
\par
Thus we demonstrate that an increase in the average number of photons $N$ results
in corresponding increase of the time interval for the applicability of our
description of squeezing dynamics. On the other hand, the number of photons
$N\geq 10^7$ initially pumped to the system is quite realistic for
contemporary experiments on light squeezing \cite{21,22,23,24}.
\par
Let us now discuss the stability of chaotic squeezing. The time-dependence
of the optimal local oscillator phase $\theta_{min}$ given in
(\ref{28}) for the cases of
regular and mild chaos is shown in Fig. 7  for two slightly different initial
conditions $z_0=1$ and $z_0=1.001$. The deviation of the optimal local oscillator
phase values related to different $z_0$ in the case of regular dynamics is
small during all time evolution presented in Fig.7
(see dashed and dotted curves). For chaos ($T=7$, $\varepsilon=0.1$), the
deviation is sufficient only at the fifth kick (boxes and crosses in Fig.7).
Thus, for not very strong chaos, there is a time interval during which squeezing
is enhanced and still stable.
\section{Conclusion}
In Ref. \cite{24} a fundamental question was asked: ``What is the
best squeezing device?'' The  answer founded in \cite{24} is
based on the fact that ``the quantum
noise reduction is effective below and above the bifurcation points between
dynamical regimes''. The reason for enhanced squeezing consists in the fact
that ``the same classical linearized equations are used
to study the stability of the
system and to calculate its quantum fluctuations'' \cite{24}. It should be
stressed that only integrable or near integrable systems with regular dynamics
were discussed in \cite{24}.
The results of our paper demonstrate that in addition to a very narrow class of
integrable systems near the threshold of instability, there is another wide
class of potentially ``the best squeezing devices'': systems at the
transition to quantum
chaos operating during the time interval of the well-defined quantum-classical
correspondence.
\par
In this paper we have presented a semiclassical theory for the quantum
systems with one
and half degrees of freedom and illustrated our findings on the model of
kicked nonlinear oscillator.
Direct numerical observation of enhanced squeezing  in the model of a kicked
quantum rotator at the transition to quantum chaos (nonperturbative approach)
will be presented elsewhere \cite{51}.
\acknowledgements
We would like to thank  Boris Chirikov,
Claude Fabre, Zden\v{e}k Hradil,
Anton\'{\i}n Luk\v{s}, and Vlasta Pe\v{r}inov\'{a}
for discussions. Part of the work was done during the visit of K. N. A.
to the Department of Physics, University of Illinois at Urbana-Champaign.
K. N. A. thanks Prof. David Campbell for hospitality.
K. N. A. also thanks Prof. Juhani Kurkij\"{a}rvi for hospitality in \AA bo.
This research was partially supported by Russian Fund for Basic Research
(96-02-16564), INTAS (94-2058),
Academy of Finland (35874), Krasnoyarsk Regional Science Foundation,
Czech Grant Agency (202/96/0421)
and Czech Ministry of Education (VS96028).

\appendix
\section{}

In this appendix we briefly describe the derivation of the coupled maps (\ref{32}),
and (\ref{33}) for the mean values and second order cumulants.
\par
Begin with the equations for mean values (\ref{32a}). Substituting the
Hamiltonian
of the nonlinear oscillator (\ref{5}) with $F(t)$ given in
(\ref{31'}) to (\ref{20a})
and using the smallness of quantum correction $N^{-1} Q$, we first neglect
the influence of the quantum correction and find the solution of the classical equations
of motion given by the map (\ref{31}) and then substitute this classical
solution into the expression for the quantum correction
$Q(z\rightarrow z_{cl}, z^*\rightarrow z^*_{cl})$. Such a perturbative approach
is valid during the time interval (\ref{24})
of the well-defined quantum-classical correspondence
for chaotic systems, and the time interval (\ref{25}) for the systems
with regular dynamics.
\par
To find the explicit form of the quantum correction $Q$, we remember that it has
the form of the second differential of $z(t)$ (see (\ref{12}) and (\ref{20b})).
Determining $d^2 z_{n+1}$, we obtain the expression (\ref{33})
for the quantum correction $Q$.
\par
We now turn to the derivation of the maps (\ref{32b}) and (\ref{32c}) for cumulants
$B$ and $C$.
As it has been shown in Sec. 2,
the equations of motion for $B$ and $C$ may
be obtained by linearization of classical motion equations near
the classical trajectory taking into account the quantum
correction, and by using substitution
$B\rightarrow |dz|^2$, $C\rightarrow (dz)^2$.
We adopt this finding for the derivation of maps for $B$ and $C$.
Utilizing again the perturbative approach, we first find the differential
of the map
(\ref{32a}) neglecting the influence of the quantum correction
\begin{equation}
\label{A1}
D z_{n+1}=e^{-i T \left( \Delta+ \mid z_n -i\varepsilon\mid^2\right)}
\{-i T (z_n-i\varepsilon)\left[ (z_n-i\varepsilon) d z_n^* +
(z_n^*+i\varepsilon) d z_n \right] +d z_n \}.
\end{equation}
From (\ref{A1}) we obtain the maps for quadratic variables
$B_n\equiv |d z_n|^2$ and
$C_n\equiv( d z_n)^2$ in the forms (\ref{32b}) and (\ref{32c}). Finally,
we suppose that the
variables $z_n$,
$z^*_n$ included in Eqs. (\ref{32b}) and (\ref{32c}) are calculated with account of
quantum correction, i. e., using
formula (\ref{32a}). The perturbative approach used in the derivation
of (\ref{32b}) and (\ref{32c}) is valid during time interval of well-defined
quantum-classical correspondence when the quantum corrections to the classical
equations are small. To avoid misprints we checked the expressions
for our coupled maps (\ref{32}) and (\ref{33}) by symbolic computation
in the package MATHEMATICA.


\begin{figure}
\caption{Nonlinear dynamics of kicked classical nonlinear oscillator
($N \rightarrow \infty$):
phase portrait
and time dependence of the intensity $|z_n|^2$ for the regular behavior
at $T=3$ (a, b), for weak chaos at $T=6.4$
(c, d), and  for hard chaos at $T=10$ (e, f). The initial condition is $z_0=1$,
and $\varepsilon=0.1$, $\Delta=1$.
Time is measured in the number of kicks.}
\label{fig1}
\end{figure}
%

\begin{figure}
\caption{
Time dependence of the principal squeezing (a), of the distance
between two initially
closed trajectories (b), and of the quantum correction to the motion
of wave packet center (c).
Curve 1 corresponds to
the regular dynamics at $T=3$, curve 2 to the mild chaos ($T=7$),
and curve 3 to the hard chaos ($T=10$). The average photon number
is $N=10^9$,
The initial condition and other parameters are the same as in Fig. 1. }
\label{fig2}
\end{figure}
%

\begin{figure}
\caption{
In the upper part of the figure, the logarithm of the distance between two
initially closed classical trajectories ${\log}_{10} d$
after seven kicks is plotted
as a function  of kicking period $T$ defined in dimensionless units
(see main text).
In the lower part, the minimum
of the logarithm of the principal squeezing $\min{\log}_{10} S$ during seven kicks
as a function of the kick period $T$ is plotted for the average photon
number $N=10^9$.
In the center of the figure, big intervals of $T$ with primary regular and
primary chaotic behavior are marked. The amplitude of perturbation
and detuning are fixed: $\varepsilon=0.1$, $\Delta=1$. The initial condition
is the same as in Fig.1. }
\label{fig3}
\end{figure}
%

\begin{figure}
\caption{Local Lyapunov rate ${\log}_{10} d$ after nine kicks and
the minimum of principal squeezing during nine kicks as
functions of dimensionless perturbation
amplitude $\varepsilon$ at fixed $T=1.$
Other parameters are the same as in Fig. 3.}
\label{fig4}
\end{figure}
%

\begin{figure}
\caption{(a) Same as in Fig. 4 but after only three kicks instead of nine.
Point 1 corresponds to $\varepsilon=1.8$ and point 2 to
$\varepsilon=2.22$.
(b) Time dependence of the classical local Lyapunov rate. Curves 1 and 2
are plotted for the same values of $\varepsilon$ as in (a) and at $T=1.$  }
\label{fig5}
\end{figure}
%

\begin{figure}
\caption{Time dependence of the principal squeezing  for chaotic motion
($T=7$, $\varepsilon=0.1$). Curve 1 corresponds to the classical limit
$N\rightarrow\infty$, and curve 2 to $N=10^7$. }
\label{fig6}
\end{figure}
%

\begin{figure}
\caption{
Time dependence of the optimal local oscillator phase $\theta_{min}$
in the case of mild chaos at $T=7$, $\varepsilon=0.1$, and for different
initial conditions: $z_0=1$ (boxes), $z_0=1.001$ (crosses). Compare with
time dependence of $\theta_{min}$ for regular motion at  $T=3$ and
$\varepsilon=0.1$ showed by dashed line for $z_0=1$ and by dotted line
for $z_0=1.001$.}
\label{fig7}
\end{figure}

\begin{references}
\bibitem{25}
To our knowledge, enhanced light squeezing at the transition to quantum
chaos was first predicted in \cite{26}.
Enhanced light squeezing at the transition to chaos in the generalized Tavis-Cummings
model was studied in detail in \cite{28}. Strong squeezing of wave packets
on a time scale of well-defined quantum-classical correspondence has been
also discussed by B. V. Chirikov as characteristic of quantum chaos \cite{29}
(see also sec. 3.1 in review \cite{15}). Our present consideration of the
problem is extended variant of \cite{30}.
\bibitem{37}
A very nice discussion of strong squeezing near bifurcation points is
presented in Ref. \cite{24}, sec. 3.2.1., see also references cited therein.
\bibitem{46}
Strictly speaking, the problem of second harmonic generation does not belong
to the class of systems (\ref{3}). We analyze in this paper single mode
quantum systems with an explicit time-dependence.
However, the examples of second-harmonic
generation and optical 3-wave mixing problems demonstrate that the extension of
our formalism to
the case of multi-mode quantum models is straightforward, at least for the
case of simple power wise nonlinearity.
\end{references}
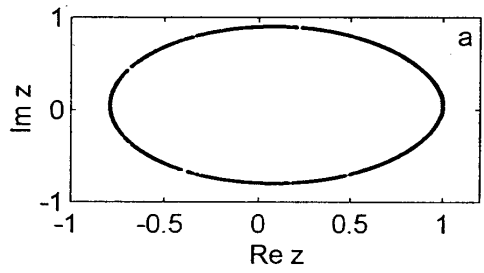
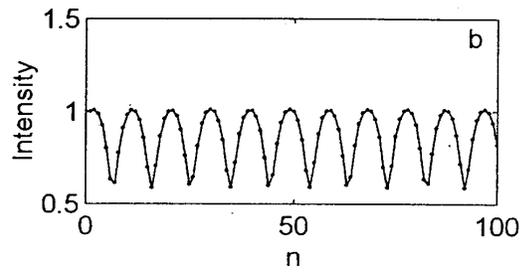
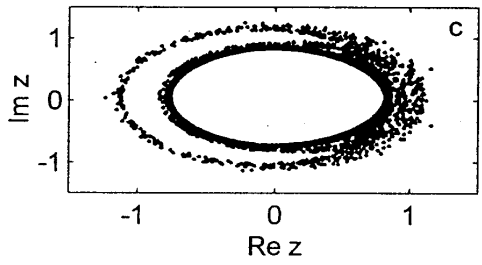
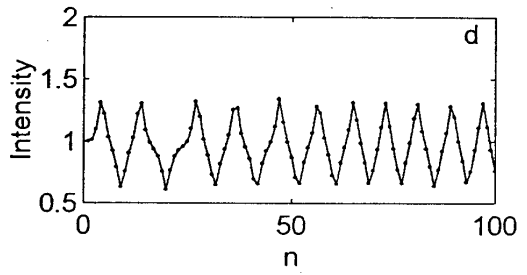
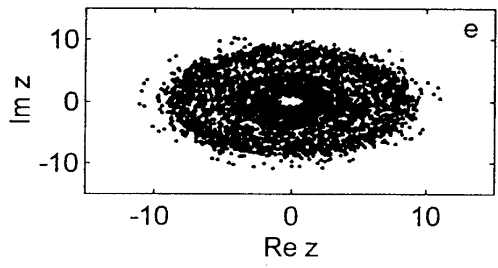
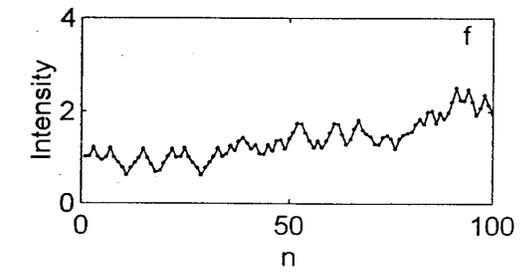
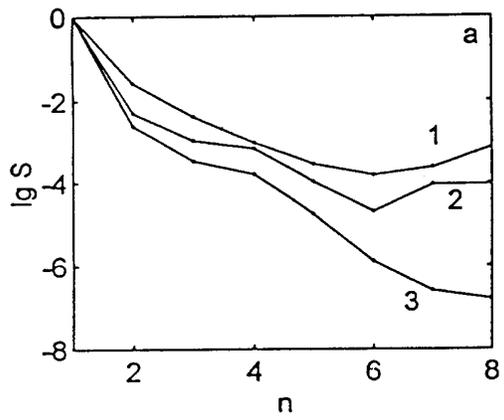
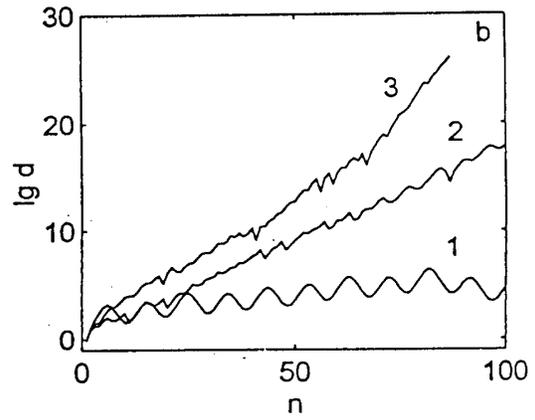
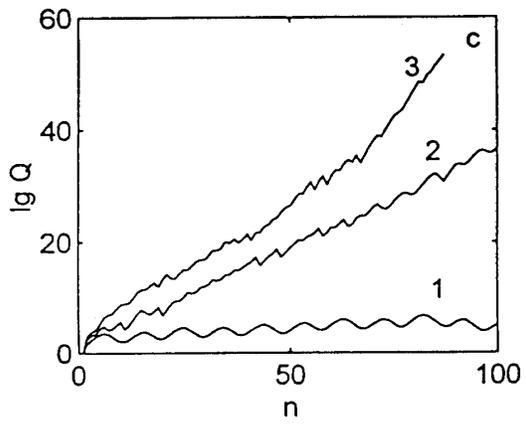
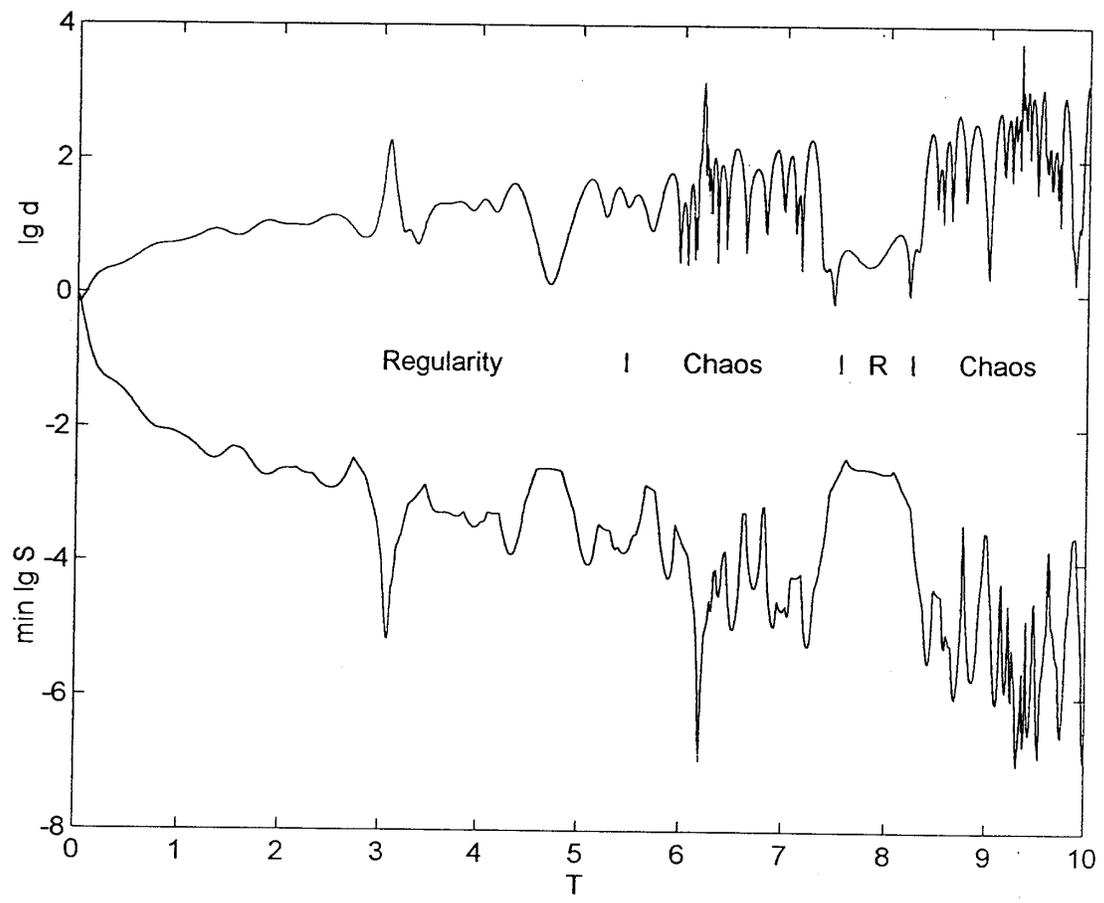
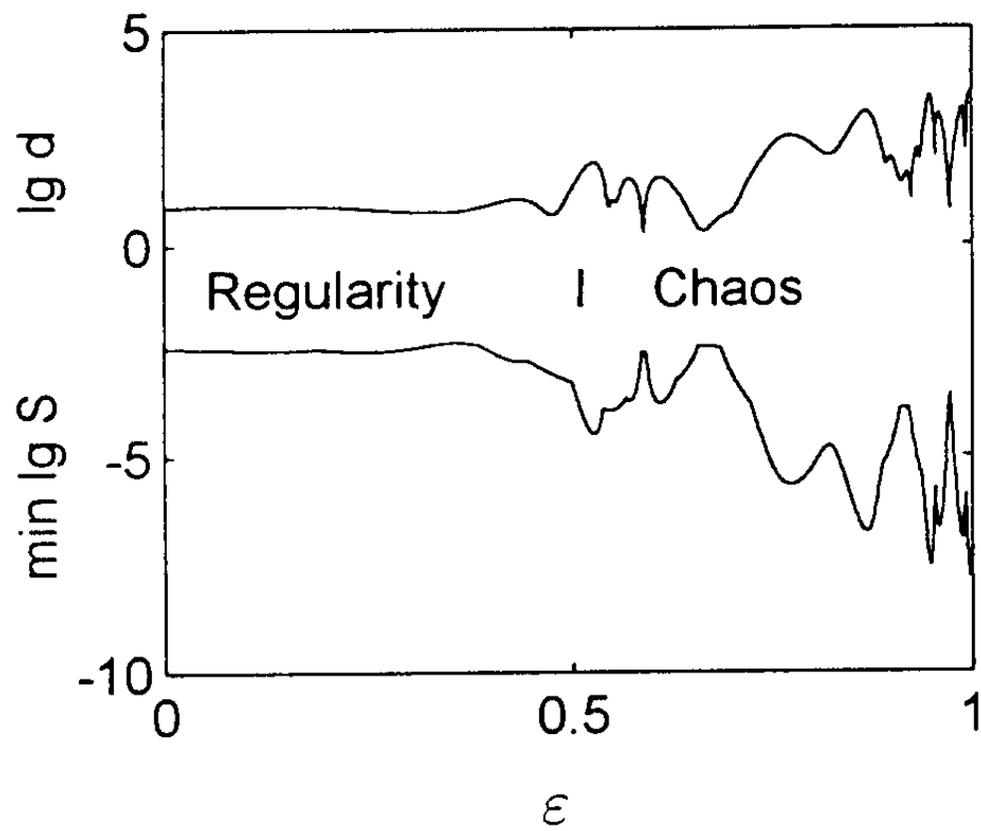
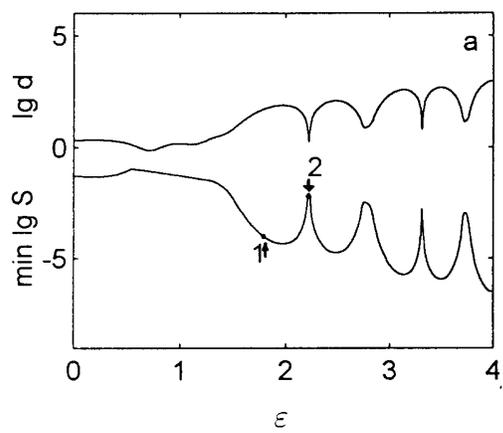
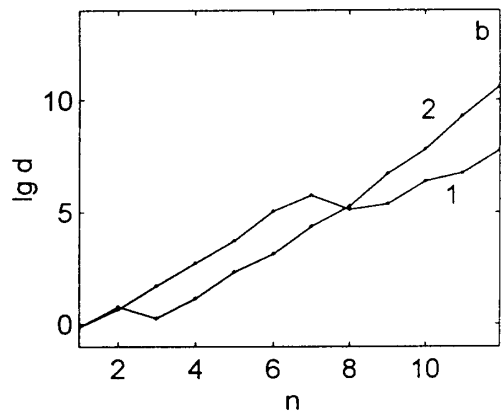
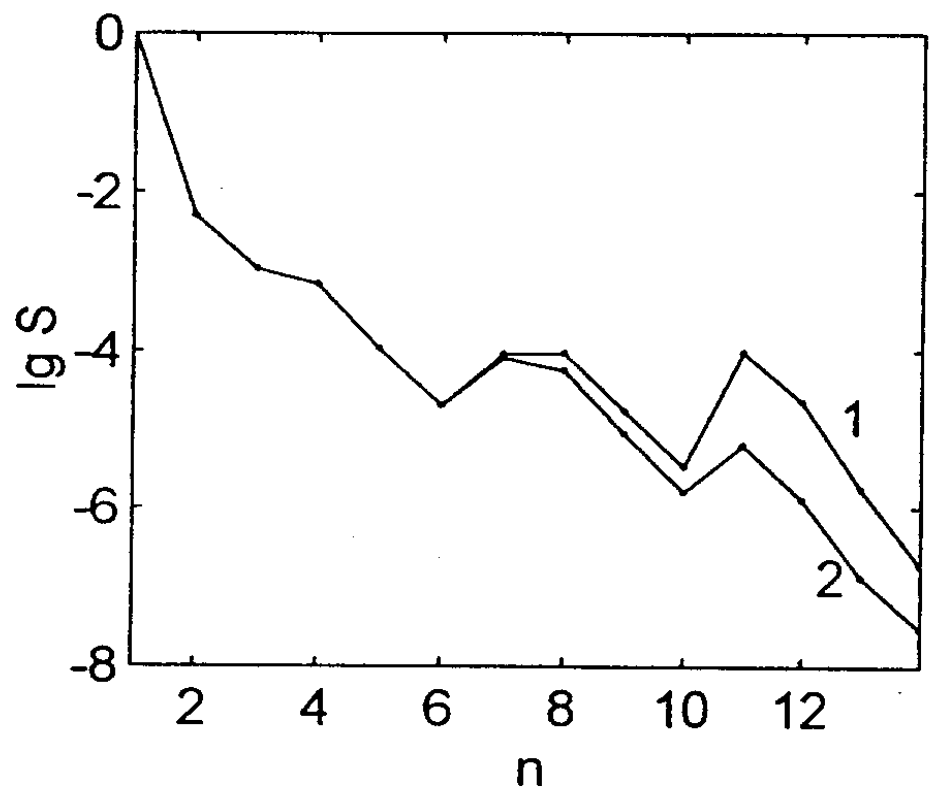
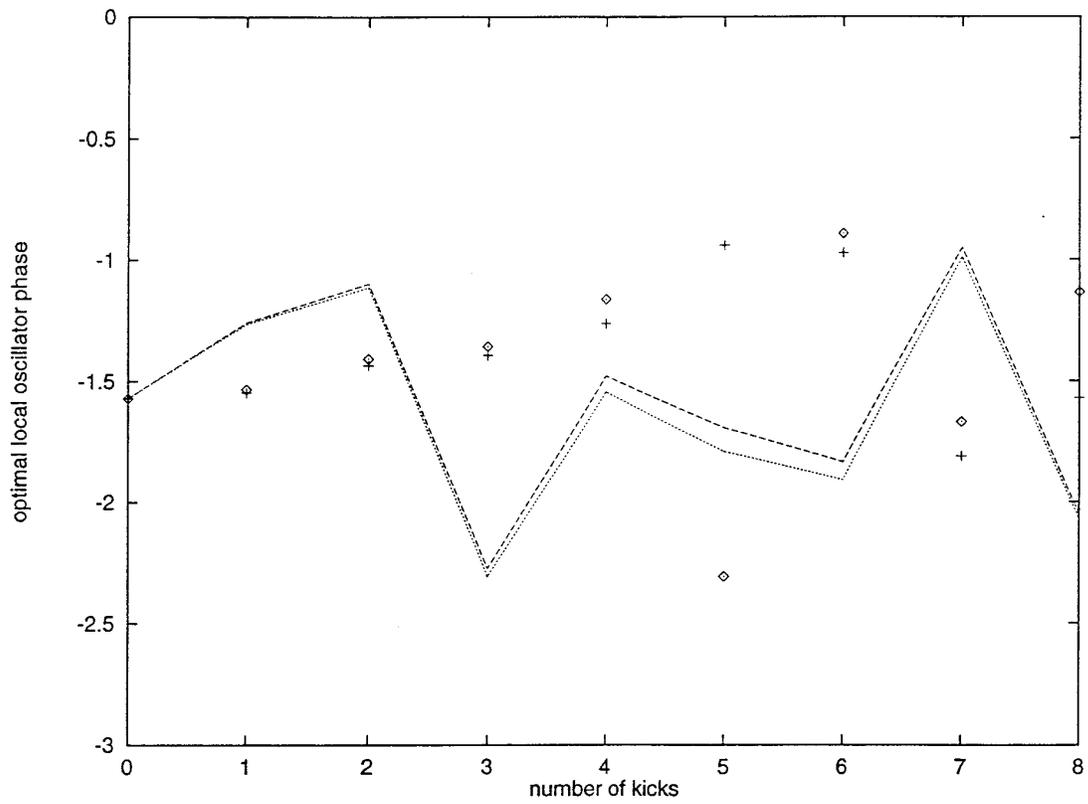
\end{document}